\documentclass{elsart} 
\usepackage{graphicx}
\usepackage{amssymb}
\usepackage[colorlinks=true,citecolor=blue,linkcolor=blue,pdftitle={Thomson scattering from
  near-solid density plasmas using soft x-ray free electron lasers},pdfauthor={Arne Hoell}]{hyperref}
\begin{document}

\begin{frontmatter}


\title{Thomson scattering from near--solid density plasmas using soft
  x--ray free electron lasers}

\author[Rostock]{A.~H\"oll}
\ead{arne.hoell@uni-rostock.de}
\author[Rostock]{Th.~Bornath}
\author[FSU]{L.~Cao}
\author[Rostock]{T.~D\"oppner}
\author[DESY]{S.~D\"usterer}
\author[FSU]{E.~F\"orster}
\author[Rostock]{C.~Fortmann}
\author[LLNL]{S.H.~Glenzer}
\author[RAL]{G.~Gregori}
\author[MBI]{T.~Laarmann}
\author[Rostock]{K.-H.~Meiwes-Broer}
\author[Rostock]{A.~Przystawik}
\author[DESY]{P.~Radcliffe}
\author[Rostock]{R.~Redmer}
\author[Rostock]{H.~Reinholz}
\author[Rostock]{G.~R\"opke}
\author[Rostock]{R.~Thiele}
\author[Rostock]{J.~Tiggesb\"aumker}
\author[DESY]{S.~Toleikis}
\author[Rostock]{N.X.~Truong}
\author[DESY]{T.~Tschentscher}
\author[FSU]{I.~Uschmann}
\author[FSU]{U.~Zastrau}
\address[Rostock]{Institut f\"ur Physik, Universit\"at Rostock,
  Universit\"atsplatz 3, 18051 Rostock, Germany}
\address[FSU]{Institut f\"ur Optik und Quantenelektronik,
  Friedrich-Schiller-Universit\"at Jena
  Max-Wien-Platz 1,
  07743 Jena, Germany}
\address[DESY]{Deutsches Elektron-Synchrotron DESY, Notkestr. 85,
  22607 Hamburg, Germany}
\address[LLNL]{L-399, Lawrence Livermore National Laboratory,
  University of California, P.O. Box 808, Livermore, CA 94551, USA}
\address[RAL]{CCLRC, Rutherford Appleton Laboratory, Chilton, Didcot
  OX11 0QX, Great Britain and Clarendon Laboratory, University of Oxford,
  Parks Road, Oxford, OX1 3PU, Great Britain}
\address[MBI]{Max-Born-Institut, Max-Born-Strasse 2a, 12489 Berlin, Germany}
\begin{abstract}
  We propose a collective Thomson scattering experiment at the VUV
  free electron laser facility at DESY (FLASH) which aims to diagnose
  warm dense matter at near--solid density. The plasma region of
  interest marks the transition from an ideal plasma to a correlated
  and degenerate many--particle system and is of current interest,
  e.g. in ICF experiments or laboratory astrophysics. Plasma
  diagnostic of such plasmas is a longstanding issue. The collective
  electron plasma mode (plasmon) is revealed in a pump--probe
  scattering experiment using the high--brilliant radiation to probe
  the plasma.  The distinctive scattering features allow to infer
  basic plasma properties. For plasmas in thermal equilibrium the
  electron density and temperature is determined from scattering off
  the plasmon mode.
\end{abstract}

\end{frontmatter}

\section{Introduction}
\label{sec:intro}
Accurate measurements of plasma temperatures and densities are
important for understanding and modeling contemporary plasma
experiments in the warm dense matter (WDM) regime\,\cite{Lee03}. WDM
is characterized by a free electron density of $n_{e}=10^{21} -
10^{26}\,{\rm cm}^{-3}$ and temperatures of several eV. In this regime
bound and free electrons become strongly correlated and medium and
long--range order are built up.  Of special interest is WDM at
near--solid density ($n_{e}=10^{21} - 10^{22} {\rm cm}^{-3}$, $T_{e}=1
- 20\,{\rm eV}$) where the transition from an ideal plasma to a
degenerate, strongly coupled plasma occurs. In particular, transient
plasma behavior at these conditions is observed in dynamical
experiments, like laser shock--wave or Z--pinch experiments and are of
particular importance for inertial confinement fusion (ICF)
experiments. In such applications the plasma evolution follows its
path through the largely unknown domain of WDM. Further applications
of this plasma domain are found, e.g., in high--energy density
physics, astrophysics and material sciences.

A rigorous understanding of highly correlated plasmas is a
long--standing and presently unresolved problem. Only recently, with
the availability of high--brilliant and coherent VUV and x-ray
radiation at a frequency larger than the density dependent electronic
plasma frequency $\omega_{pe}$, it became feasible to penetrate
through dense plasmas and study basic plasma properties from scattered
spectra\,\cite{Riley00,Landen01,Gregori03,Hoell04,Redmer05}.  The
recent effort to develop diagnostic methods using Thomson scattering
is an important first step towards a systematic understanding of WDM.
Available radiation sources to probe such plasmas are backlighter
systems in the x-ray regime as well as free electron laser (FEL)
radiation currently available in the VUV.  Backlighters have been
developed in ICF related research\,\cite{Landen01_2} and have been
applied to solid density plasmas. The first experiment to measure the
spectrally resolved x-ray Thomson scattering spectrum in solid density
Be plasma has been reported in\,\cite{Glenzer03}. Using the
$4.75\,{\rm keV}$ titanium He-$\alpha$ backlighter, the
non--collective Thomson scattering spectrum from the thermally
distributed electrons allowed the measurement of the Compton--shifted
electron distribution function, which was used to determine the plasma
density, temperature and ionization degree. A pioneering scattering
experiment from the collective electron plasma mode (plasmon) at solid
density using a Cl~Ly-$\alpha$ backlighter at $2.96\,{\rm keV}$ has
been performed recently\,\cite{Glenzer06}.  A new type of radiation
sources for WDM studies became available with the advances of FELs.
For instance, the FLASH facility at DESY, has successfully started
user experiments in the VUV range $13 - 50\,{\rm
  nm}$\,\cite{Ayvazyan06,Stojanovic06,Chapman06} using the
SASE\footnote{SASE: self amplification of stimulated emission}
principle\,\cite{Saldin80,Bonifacio84}.  A proof--of--principle
collective Thomson scattering experiment at FLASH ($25\,{\rm nm}$)
will be performed in March 2007.  This experiment aims to demonstrate
Thomson scattering with FEL radiation at near--solid density plasmas
as a diagnostic method which is described in this paper in detail.
With further advance in FEL technology with respect to bandwidth,
photon energy and brilliance, the method presented here can be
developed to a standard diagnostic tool for WDM research.

Scattering from ideal plasmas has much been
studied\,\cite{Hughes75,Sheffield75,Bekefi66} and can be described
theoretically within the random phase approximation (RPA). It has been
shown\,\cite{Hoell04,Redmer05} that in the region of near--solid
density collisions significantly modify the scattering spectrum and a
theory beyond the RPA is needed. The elaboration of such theories of
non--ideal plasmas\,\cite{Klimontovich82} is more complicated and
different ways have been followed in the past. Local field
corrections\,\cite{Gregori04}, incorporated by
Ichimaru\,\cite{Ichimaru85,Ichimaru94} using a general
density--response formalism, consistently improve the RPA.
Alternatively, a theory based on the dielectric function which can be
expressed in terms of correlation functions has been studied. For
their calculation perturbative methods have been
developed\,\cite{Roepke98,Roepke98_2}.  Applications for optical and
transport properties in WDM are
studied\,\cite{Wierling01,Reinholz05,Dharma-wardana06}.

The paper is organized as follows. In section~\ref{sec:Theory} we
outline the theoretical basis for Thomson scattering in WDM, introduce
the dynamical structure factor as a central quantity and describe its
calculation. Results of the calculations and estimations are used to
discuss the applicability of collective Thomson scattering for plasma
diagnostics in section~\ref{sec:Results}. In
section~\ref{sec:Experiment} we describe the planned experiment at
FLASH.  Based on the results of section~\ref{sec:Results}, we specify
out requirements and restrictions relevant for the experiment.  In
section~\ref{sec:Summary} we give a summary and address future
extensions of the method.

\section{Theory}
\label{sec:Theory}
\subsection{Scattering Geometry}
\label{subsec:ScattGeometry}
\begin{figure}[ht]
\begin{center}
\includegraphics[width=0.6\textwidth,angle=0]{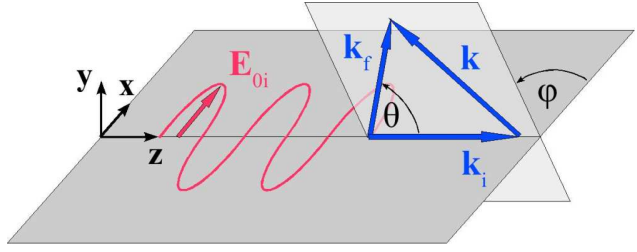}
\caption{\label{fig:PolarizationScheme}Scattering geometry}
\end{center}
\end{figure}

In this section we will outline the theory needed to describe the
scattering of coherent radiation from an equilibrium plasma at
near--solid density.  The scattering geometry is shown in
Fig.~\ref{fig:PolarizationScheme}.  The plasma is irradiated by the
linearly polarized FEL probe--beam in $z$--direction with the
polarization pointing into the $x$--direction.  The detector is
located in the direction of the scattered wave vector ${\bf k}_{f}$ at
the distance $R$ much larger than the plasma extension. The direction
of ${\bf k}_{f}$ is characterized by the scattering angle $\theta$ and
the azimuthal angle $\varphi$ as shown in
Fig.~\ref{fig:PolarizationScheme}.
The momentum transfer of the scattered photon is given by ${\bf
  k}={\bf k}_{f}-{\bf k}_{i}$ and its energy transfer by
$\omega=\omega_{f}-\omega_{i}$. The modulus of the initial and final
wave vector is given by
\begin{eqnarray}
\label{eq:Modk_i}
& &
k_{i}=|{\bf k}_{i}| 
= \frac{\omega_{i}}{c}\sqrt{1-\frac{\omega_{pe}^{2}}{\omega_{i}^{2}}}
\\
& &
\label{eq:Modk_f}
k_{f}=|{\bf k}_{f}| 
= \frac{\omega_{f}}{c}\sqrt{1-\frac{\omega_{pe}^{2}}{\omega_{f}^{2}}}
\approx \frac{\omega_{f}}{c}\sqrt{1-\frac{\omega_{pe}^{2}}{\omega_{i}^{2}}} ~.
\end{eqnarray}
The square root terms in Eq.~(\ref{eq:Modk_i}) and (\ref{eq:Modk_f})
account for the dispersion of incoming and scattered wave. 
In the density region of $10^{21}-10^{22}\,{\rm cm}^{-3}$ the electron
plasma frequency is of the order of
\begin{math}
\omega_{pe}=\sqrt{n_{e}e^{2}/\epsilon_{0}m_{e}}
\approx 1-4\,{\rm eV}
\end{math}
and the probe radiation in the VUV--region is of the order of
$\omega_{i}\approx\omega_{f}=50-100\,{\rm eV}$. Therefore, we have
\begin{math}
|\omega|\lesssim\omega_{pe}\ll\omega_{i}\approx\omega_{f},~
\end{math}
and the approximation in Eq.~(\ref{eq:Modk_f}) is well justified.
From the scattering geometry given in
Fig.~\ref{fig:PolarizationScheme} one finds for the modulus of the
momentum transfer
\begin{math}
k = \sqrt{k_{i}^{2}+k_{f}^{2} - 2 k_{i}k_{f}\cos\theta} 
\end{math}
and using Eqs.~(\ref{eq:Modk_i}) and (\ref{eq:Modk_f})
\begin{eqnarray}
\label{eq:MomTransfer}
k = \frac{4\pi}{\lambda_{0}}\sin(\theta/2)
\,\sqrt{1 + \frac{\omega}{\omega_{i}} 
+ \frac{\omega^{2}}{\omega_{i}^{2}}\,\frac{1}{4\sin^{2}(\theta/2)}} 
\;\sqrt{1-\frac{\omega_{pe}^{2}}{\omega_{i}^{2}}}~,\quad
\omega_{i} = \frac{2\pi c}{\lambda_{0}} ~.
\end{eqnarray}
As pointed out, $|\omega|/\omega_{i}$ and $\omega_{pe}/\omega_{i}$ are
small quantities, and it is easily observed from
Eq.~(\ref{eq:MomTransfer}) that the momentum transfer $k$ is well
approximated by the relation $k =
\frac{4\pi}{\lambda_{0}}\sin(\theta/2)$ for elastic scattering. This,
however, is only valid for $\theta > |\omega|/\omega_{i}$, a
condition fulfilled in most scattering experiment.  

\begin{figure}[ht]
\begin{center}
\centerline{%
{\footnotesize
$\quad\left| \widehat{\mathbf{k}}_{f} \times \left( \widehat{\mathbf{k}}_{f} 
\times \widehat{\mathbf{E}}_{0i}^{}\right)\right|^{2}$}}
\vspace*{-0.5ex}
\centerline{%
\includegraphics[width=0.6\textwidth,angle=0]{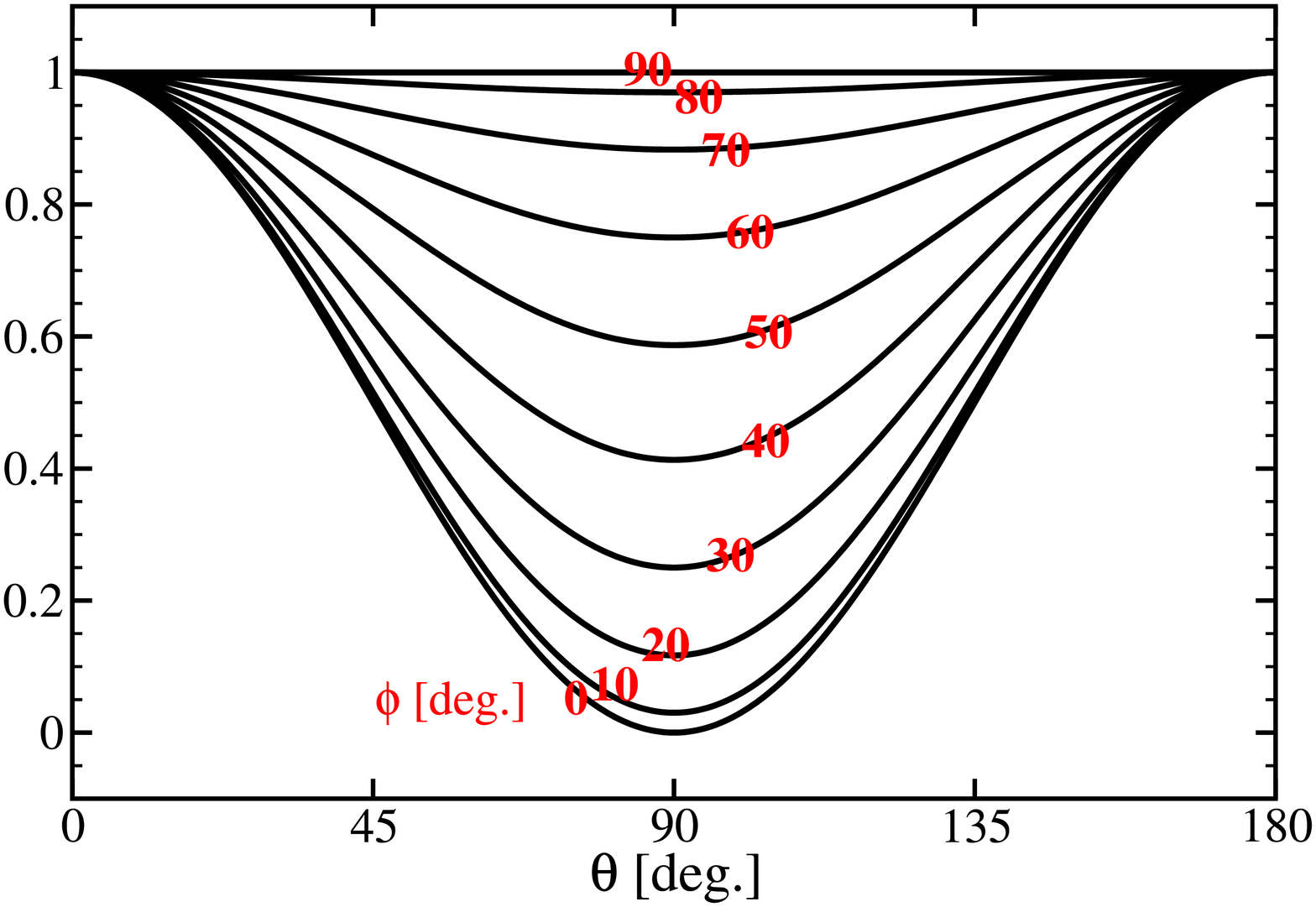}}
\caption{\label{fig:PolarizationDependence}Dependence of the scattered
  light on laser
  polarization, scattering angle $\theta$ and azimuthal angle $\varphi$.}
\end{center}
\end{figure}

The scattered power $P_s$ from the electrons into a frequency interval
$d\omega$ and solid angle $d\Omega$ is given by\,\cite{Sheffield75}
\begin{eqnarray}
\label{eq:ScattPower}
P_{s}(\mathbf{R},\omega_{}) d\Omega d\omega_{} =
\frac{P_{i}r_{0}^{2}d\Omega}{2\pi A} \left|
\widehat{\mathbf{k}}_{f}\times (\widehat{\mathbf{k}}_{f} \times
\widehat{\mathbf{E}}_{0i}) \right|^{2} N_{}^{} S(\mathbf{k},\omega_{})
d\omega_{}~.
\end{eqnarray}
In Eq.~(\ref{eq:ScattPower}) $P_{i}$ denotes the incident FEL power,
$r_{0}=e^2/m_{e}c^{2}=2.8\times10^{-15}\,{\rm m}$ the classical
electron radius, $A$ the plasma area irradiated by the FEL, $N_{}^{}$
the number of nuclei and $S({\bf k}, \omega)$ the total electron
dynamical structure factor.  The polarization term
\begin{math}
|\widehat{\mathbf{k}}_{f}\times (\widehat{\mathbf{k}}_{f} \times
\widehat{\mathbf{E}}_{0i})|^{2}
\end{math}
reflects the dependence of the scattered power on the incident laser
polarization with the hat denoting unit vectors. For linear polarization
and for unpolarized light, respectively, this term is given by
\begin{eqnarray}
\label{eq:Polarisation}
\left| \widehat{\mathbf{k}}_{f} \times \left( \widehat{\mathbf{k}}_{f} 
\times \widehat{\mathbf{E}}_{0i}^{}\right)\right|^{2} 
= \left\{
\begin{array}{ll}
(1 - \sin^{2}\theta \cos^{2}\varphi )
&\mbox{lin. polarized}
\\[1ex]
(1 - \frac{1}{2}\sin^{2}\theta )
=\frac{1}{2}\left( 1+\cos^{2}\theta \right)
&\mbox{unpolarized}
\end{array}
\right.
\end{eqnarray}
In the case of linear polarization the dependence on the scattering
angle $\theta$ is shown in Fig.~\ref{fig:PolarizationDependence}. The
strongest dependence on $\theta$ is observed for $\varphi=0^{\circ}$,
whereas at $\varphi=90^{\circ}$ the polarization term is independent
of the scattering angle $\theta$.
As seen from Eq.~(\ref{eq:ScattPower}), the scattered power depends on
the setup of the scattering experiment (initial probe power,
scattering angle, probe wavelength, probe polarization), the density
and plasma length, and on the total electron dynamical structure
factor $S({\bf k}, \omega)$. The dynamical structure factor is the
Fourier transform of the electron--electron density fluctuations.
It is a central quantity since it contains the details of the
correlated many--particle system\,\cite{Ichimaru94,Ichimaru88}. The definition
and calculation of $S({\bf k}, \omega)$ are given in
section~\ref{subsec:scattering}.

\subsection{Thomson Scattering in WDM}
\label{subsec:scattering}
A characterization of plasmas in equilibrium at different densities
and temperatures can be performed by simple estimations of dimensionless
parameters. The coupling parameter $\Gamma$ is defined as the ratio of
the Coulomb energy between two charged particles at a mean particle
distance $\bar{d}$ and the thermal energy. Using the
Wigner--Seitz radius for $\bar{d}$, we find an expression for
the electron coupling parameter
\begin{eqnarray}
\label{eq:Gamma}
\Gamma = \frac{e^{2}}{4\pi\varepsilon_{0}\bar{d} k_{B}T_{e}}
\quad,\quad
\bar{d} = \left( \frac{4\pi n_{e}}{3} \right)^{-1/3} ~.
\end{eqnarray}
If $\Gamma<1$ the plasma is weakly coupled, and $\Gamma\ll 1$ denotes
the ideal plasma regime. Correlations become more important in the
coupled plasma regime, where $\Gamma\gtrsim 1$.  
The degeneracy parameter $\Theta$ estimates the role of quantum statistical
effects in the system and is given by the ratio of the thermal energy
and the Fermi energy $E_{F}$
\begin{eqnarray}
\label{eq:Theta}
\Theta = \frac{k_{B}T_{e}}{E_{F}}
\quad,\quad
E_{F} = \frac{\hbar^{2}}{2m_{e}}\left( 3\pi^{2} n_{e} \right)^{2/3} ~.
\end{eqnarray}
In a degenerate plasma, the Fermi energy is larger than the thermal
energy, i.e. $\Theta<1$, and most electrons populate states inside the
Fermi sea where quantum effects are of importance. Contrary to that,
for $\Theta>1$ the role of quantum effects decreases.

The dimensionless scattering parameter $\alpha$ compares the length
scale of electron density fluctuations $\ell\approx 2\pi/k$ measured
in the scattering experiment to the screening length $\lambda_{sc}$ in
the plasma. The scattering parameter is defined as
\begin{eqnarray}
\label{eq:alpha}
\alpha = \frac{1}{k\lambda_{sc}}
\quad,\quad
\lambda_{sc}^{-2} \to \lambda_{sc,e}^{-2} = \kappa_{sc,e}^{2} 
= \frac{e^2m_e^{3/2}}{\sqrt{2}\pi^{2}\epsilon_{0}\hbar^{3}}
\int_0^{\infty}dE\,E^{-1/2}f^{e}(E) ~,
\end{eqnarray}
where we used the electron screening length $\lambda_{sc,e}$ which is
written in terms of the Fermi--Dirac distribution function $f^{e}(E)$
of the electrons and accounts for quantum effects\,\cite{Reinholz05}.
In the case of a classical Maxwell--Boltzmann electron distribution
function, the screening length in Eq.~(\ref{eq:alpha}) reduces to the
Debye screening length $\kappa_{sc,e}^2 \to \kappa_{D,e}^2 =
n_{e}e^{2}/(\epsilon_{0}k_{B}T_{e})$.  For $\alpha>1$, i.e. in the
collective scattering regime, only density fluctuations larger than
the screening length are observed, while at $\alpha<1$, i.e. in the
non-collective scattering regime, the density fluctuations of
individual electrons are resolved.  As seen from
Eq.~(\ref{eq:MomTransfer}), $\ell=2\pi/k$ is mainly determined by the probe
wavelength and the scattering angle (i.e. by the setup of the
scattering experiment) while $\lambda_{sc,e}$ is determined by the
plasma properties like density and temperature (see
Eq.~(\ref{eq:alpha})).
\begin{figure}[ht]
\begin{center}
\includegraphics[angle=270,width=0.7\textwidth]{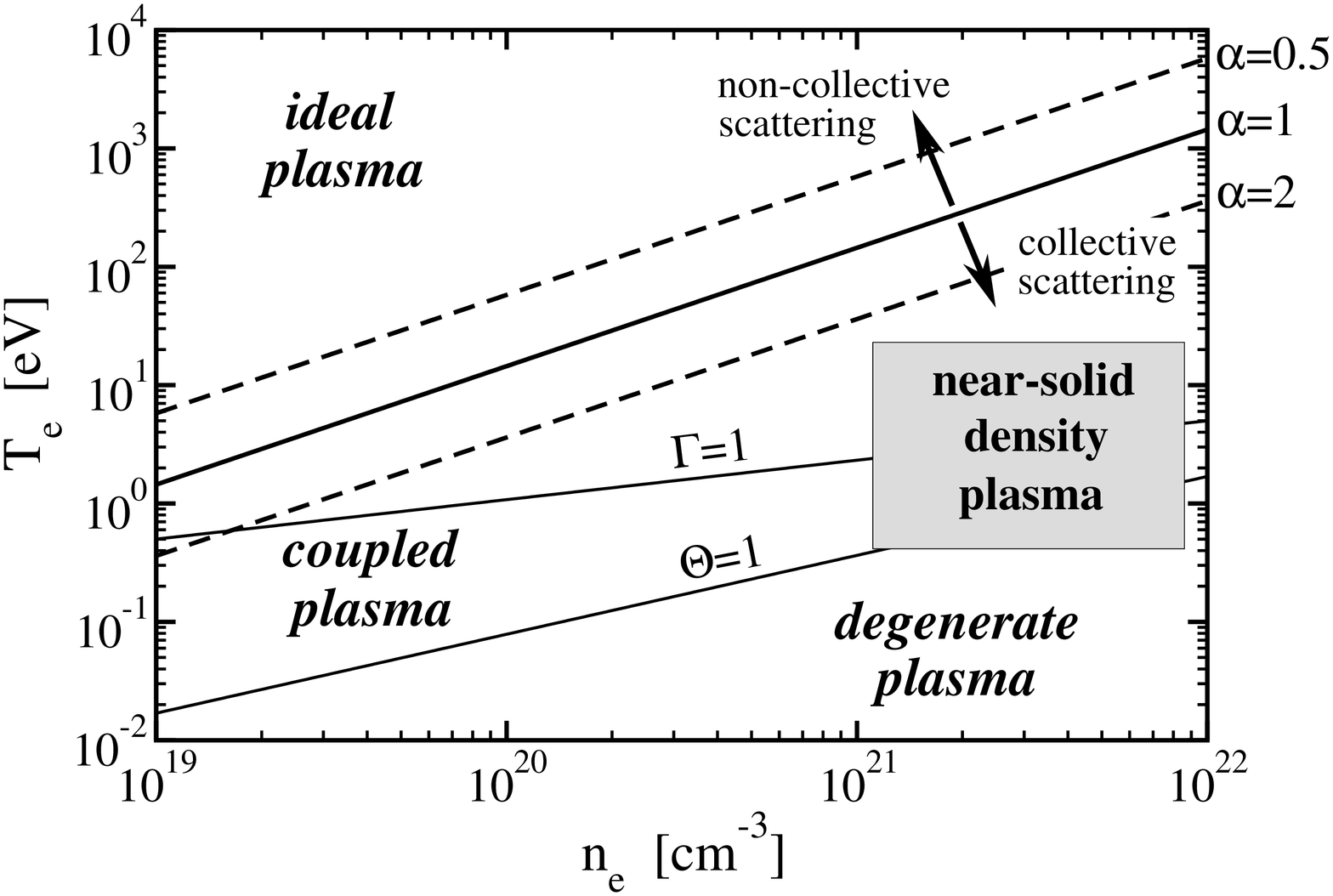}
\caption{\label{fig:nT}Plasma parameters in the density--temperature plane. The
  region of moderate temperature, near--solid density plasmas is
  shown. Also, the electron coupling parameter $\Gamma$, degeneracy
  parameter $\Theta$ as well as the scattering parameter $\alpha$ is
  shown as defined in Eqs.~(\ref{eq:Gamma}), (\ref{eq:Theta}) and
  (\ref{eq:alpha}). Assumed are the scattering angle
  $\theta=90^{\circ}$ and probe laser wavelength $\lambda_{0}=25\,{\rm
    nm}$.}
\end{center}
\end{figure}
The density--temperature plot in Fig.~\ref{fig:nT} shows different
plasma regions. In particular, the region of moderate temperature and
near--solid density (gray region), overlaps with the ideal plasma,
coupled plasma and degenerate plasma region. Depending on the
scattering geometry conditions and probing wavelength, this plasma
region is accessible applying both, collective and non--collective
scattering. At a wavelength $\lambda_{0}=25\,{\rm nm}$, however, there
is only collective scattering ($\alpha>0$) for arbitrary scattering
angle observable from this plasma region.

As outlined in conjunction with Eq.~(\ref{eq:ScattPower}), the
scattering from the plasma is determined by the mean
electron--electron density fluctuations per nucleus which is
expressed by the dynamical structure factor $S(\mathbf{k},\omega)$.
The structure factor is defined in terms of the electron 
density response function $\chi_{ee}^{tot}$ via a
fluctuation--dissipation theorem (FDT)\,\cite{Kubo66}
\begin{eqnarray}
\label{eq:FDT}
S({\bf k}, \omega) = \frac{\hbar}{n_{}^{}}\,\frac{1}{1
-{\rm e}^{-\beta\hbar\omega}}\,
{\rm Im}\,\chi_{ee}^{tot}({\bf k}, \omega) ~,
\end{eqnarray}
with $n$ the number density of the nuclei (atoms and ions).  The
density response function $\chi_{ee}^{tot}$ is given
as\,\cite{Selchow01}
\begin{eqnarray}
\label{eq:ChiTot}
\chi_{ee}^{tot}({\bf k}, \omega) 
= \Omega_{0}\,\frac{i}{\hbar} 
\int_{0}^{+\infty}d t\; {\rm e}^{i(\omega+i\eta) t}
\left\langle \left[ \delta n_{e}^{tot}({\bf k},t),
\delta n_{e}^{tot}(-{\bf k},0) \right]\right\rangle ~,
\end{eqnarray}
with $\delta n_{e}^{tot}=n_{e}^{tot}-\langle n_{e}^{tot}\rangle$ being
the density fluctuations of all electrons from the average density and
the brackets denoting the ensemble average. The normalization volume
is denoted by $\Omega_{0}$. The limit $\eta\to 0$ has to be taken
after the thermodynamic limit. The solution of Eq.~(\ref{eq:ChiTot})
accounting for bound states is a nontrivial problem. Here we follow
the idea of Chihara\,\cite{Chihara87,Chihara00} who applied a chemical
picture and decomposed the electrons into free and bound electrons of
density $n_{e}$ and $n_{e}^{b}$ respectively. With $Z_{f}$ free and
$Z_{b}$ bound electrons per nucleus, $n_{e}$ and $n_{e}^{b}$ are
related to the ion density $n_{i}$ according to $n_{e}=Z_{f}n_{i}$ and
$n_{e}^{b}=Z_{b}n_{i}$. Writing the total electron density
$n_{e}^{tot}=n_{e}+n_{e}^{b}$, we can decompose the total density
response function in Eq.~(\ref{eq:ChiTot}) according to
\begin{eqnarray}
\label{eq:ChiDecomp}
\chi_{ee}^{tot}
= \chi_{ff} + \chi_{bb} + \chi_{bf} + \chi_{fb} 
= \chi_{ff} + \chi_{bb} + 2 \chi_{bf} ~,
\end{eqnarray}
with the electron density response functions of the free--free,
bound--bound, bound--free and free-bound system $\chi_{ff}$,
$\chi_{bb}$, $\chi_{bf}$ and $\chi_{fb}$, respectively. The response
function are defined as in Eq.~(\ref{eq:ChiTot}) with the
corresponding electron densities.
We also used $\chi_{bf}=\chi_{fb}$.  Chihara showed that for a
classical system the total electron dynamical structure factor can be
written in the following decomposed form
\begin{eqnarray}
\label{eq:Chihara}
S(k,\omega) &=& Z_{f} S_{ee}^{0}(k,\omega)
+ \left|f_{I}(k)+q(k)\right|^{2} S_{ii}(k,\omega)
\nonumber\\
& & + Z_{b} \int d\omega' \widetilde{S}_{ce}(k,\omega-\omega')
S_{S}(k,\omega') ~.
\end{eqnarray}
The first term is due to the free electron fluctuations with $Z_{f}$
the number of free electrons per atom. The second term describes number
fluctuations of the weakly and tightly bound electrons, with $f_{I}(k)$
the ion form factor, $q(k)$ the electron screening cloud and
$S_{ii}$ the ion--ion structure factor. The last term in
Eq.~(\ref{eq:Chihara}) is attributed to inelastic Raman transitions of
core electrons into the continuum $\widetilde{S}_{ce}$, modulated with
the ion self--motion $S_{S}$.

Under the condition $E_{b}\gg \hbar^{2}k^{2}/(2m)$\,\cite{Gregori03},
where $E_{b}$ is the ionization potential of the bound states, the
first two terms in Eq.~(\ref{eq:Chihara}) are most important. For the
experimental conditions, specified in section~\ref{sec:Experiment},
the ionization potential for hydrogen and helium is
$E_{b}=13.6\,{\rm eV}$ and $24\,{\rm eV}$, respectively. The energy
transferred by the x--ray photons to the bound electrons is only $\sim
0.02\,{\rm eV}$.  Therefore, bound--free transitions can be neglected.

\subsection{Density Response Function}
\label{subsec:chi}
To shortly review the calculation of the electron density response
function we follow\,\cite{Selchow01}. We consider a plasma consisting
of free electrons and ions. The electron and the ion response
functions in a collision--free one--component plasma (OCP) model are
given by the well--known Lindhard expressions
\begin{eqnarray}
\label{eq:Lindhard}
\chi_{cc'}^{0}({\bf k},\omega) 
= \delta_{cc'}\frac{1}{\Omega_{0}} \sum_{\mathbf{p}}
\frac{f_{\mathbf{p}+\mathbf{k}/2}^{c} -
  f_{\mathbf{p}-\mathbf{k}/2}^{c}}%
{\Delta E_{\mathbf{p},\mathbf{k}}^{c} 
- \hbar (\omega + i \eta)}
= \chi_{c}^{0}({\bf k},\omega) \delta_{cc'} ~,
\end{eqnarray}
with
\begin{math}
\Delta E_{\mathbf{p},\mathbf{k}}^{c}
=E_{\mathbf{p}+\mathbf{k}/2}^{c} - E_{\mathbf{p}-\mathbf{k}/2}^{c}
=\hbar^{2}\mathbf{k}\cdot\mathbf{p}/m_{c},
\end{math}
and the Fermi function
\begin{math}
f_{p}^{c} = [\exp(\beta E_{p}^{c}-\beta\mu_{c})+1]^{-1}
\end{math}
with $\beta=1/(k_{B}T)$. The
pole appearing in Eq.~(\ref{eq:Lindhard}) is shifted by $\eta$ off the
real axis and the limit $\eta\to 0^{+}$ has to be taken after the
thermodynamic limit.  Eq.~(\ref{eq:Lindhard}) is written in matrix
form for the plasma species denoted by the indices $c$ and $c'$.  The
matrix is diagonal, i.e. by definition, there are no density
fluctuations between different plasma species within the OCP
description.

In RPA the plasma interacts via a screened interaction
$V_{cc'}^{sc}$ obtained self--consistently from the ``ring summation''
according to
\begin{eqnarray}
\label{eq:Vsc}
V_{cc'}^{sc}(\mathbf{k},\omega) 
= V_{cc'}(\mathbf{k}) + \sum_{d} V_{cd}(\mathbf{k})
\chi_{d}^{0}(\mathbf{k},\omega) \Omega_{0}
V_{dc'}^{sc}(\mathbf{k},\omega) ~,
\end{eqnarray}
which is solved by
\begin{eqnarray}
\label{eq:VscSolution}
V_{cc'}^{sc}(\mathbf{k},\omega) 
= \frac{V_{cc'}(\mathbf{k})}%
{1-\chi_{e}^{0}\Omega_{0} V_{ee}-\chi_{i}^{0}\Omega_{0} V_{ii}}
\end{eqnarray}
in the case of a Coulomb bare potential%
\footnote{Note, in\,\cite{Selchow01} arbitrary potentials are
  considered. The simplified expressions here are obtained by using
  $V_{ee}V_{ii}-V_{ei}V_{ie}=0$ valid for Coulomb interactions.},
$V_{cc'}(k)= e_{c}e_{c'}/(\epsilon_{0}\Omega_{0}k^{2})$,
considered in this paper only.
The screened interaction contains mutual screening contribution from both
plasma species (electrons and ions). 
The RPA response tensor is given by solving
\begin{eqnarray}
\label{eq:chiRPA}
\chi_{cc'}^{RPA}(\mathbf{k},\omega) = \chi_{c}^{0}(\mathbf{k},\omega)
\delta_{cc'} + \chi_{c}^{0}(\mathbf{k},\omega)\Omega_{0}
V_{cc'}^{sc}(\mathbf{k},\omega) \chi_{c'}^{0}(\mathbf{k},\omega)
\end{eqnarray}
with the matrix elements written as
\begin{eqnarray}
\label{eq:chi_ee}
\chi_{ee}^{RPA}({\bf k}, \omega) 
&=& \frac{\chi_{e}^{0} - \chi_{e}^{0} \Omega_{0} V_{ii}\chi_{i}^{0}}%
{1-\chi_{e}^{0}\Omega_{0} V_{ee}-\chi_{i}^{0}\Omega_{0} V_{ii}} ~,
\\
\label{eq:chi_ii}
\chi_{ei}^{RPA}({\bf k}, \omega) 
&=& \frac{\chi_{e}^{0} \Omega_{0}V_{ei}\chi_{i}^{0}}%
{1-\chi_{e}^{0}\Omega_{0}V_{ee}-\chi_{i}^{0}\Omega_{0}V_{ii}} ~.
\end{eqnarray}
The remaining two matrix elements are obtained by interchanging
$e\leftrightarrow i$.

To take into account the density fluctuations from the bound
electrons, not present in Eqs.~(\ref{eq:chi_ee}) and
(\ref{eq:chi_ii}), one has to solve the dielectric plasma response
including bound states\,\cite{Roepke79}. This goes beyond the scope of
this paper, and we follow the idea of Chihara\,\cite{Chihara00} by
considering the plasma in a chemical picture and decompose the
electrons as free, weakly bound and tightly bound electrons. From
Eq.~(\ref{eq:MomTransfer}) we find a maximal momentum transfer at
$\omega\approx\omega_{pe}$ for $\theta=180^{\circ}$.  Therefore, the
smallest length scale $\ell=2\pi/k$ at which density fluctuations can
be resolved in the scattering process is $\ell\ge 12\,{\rm nm}$ for a
probe wavelength $\lambda_{0}=25\,{\rm nm}$. The electron and ion size
is much smaller than $\ell$ and, consequently, details of the atomic
structure of the ion enter only in integrated form. This allows to
treat the bound electrons in a frozen core approximation. This implies
that $\chi_{ff}$, $\chi_{bb}$ and $\chi_{bf}$ for the free--free,
bound--bound and bound--free electron density response function
respectively can be written as
\begin{eqnarray}
\label{eq:chi_ff}
\chi_{ff}^{RPA}({\bf k}, \omega) 
&=& \chi_{ee}^{RPA}({\bf k}, \omega) ~, 
\\
\label{eq:chi_bb}
\chi_{bb}^{RPA}({\bf k}, \omega) 
&=& Z_{b}^{2} \chi_{ii}^{RPA}({\bf k}, \omega) ~, 
\\
\label{eq:chi_bf}
\chi_{bf}^{RPA}({\bf k}, \omega) 
&=& Z_{b}^{} \chi_{ie}^{RPA}({\bf k}, \omega) ~,
\end{eqnarray} 
where all density fluctuations are referred to the total number of
nuclei. From Eq.~(\ref{eq:ChiDecomp}) and
(\ref{eq:chi_ff})--(\ref{eq:chi_bf}) the total electron density
response function is written as
\begin{eqnarray}
\label{eq:chi_ee_tot}
\chi_{ee}^{tot}({\bf k}, \omega) 
= \chi_{ee}^{RPA}
+ Z_{b}^{2} \chi_{ii}^{RPA} 
+ 2 Z_{b}^{} \chi_{ie}^{RPA} ~.
\end{eqnarray}
Upon introduction of the partial dynamical structure factors following
Eq.~(\ref{eq:FDT})
\begin{eqnarray}
\label{eq:PartS}
S_{ee} 
= \frac{\mathcal{C}}{n_{e}}\,{\rm Im}\widetilde{\chi}_{ee}^{RPA}~,
\quad
S_{ii} = 
\frac{\mathcal{C}}{n_{i}}\,{\rm Im}\chi_{ii}^{RPA} ~,
\quad
S_{ei} = 
\frac{\mathcal{C}}{\sqrt{n_{e}n_{i}}}\,{\rm Im}\chi_{ei}^{RPA}
\end{eqnarray}
with $\mathcal{C}^{-1}=(1-\exp[-\beta\hbar\omega])/\hbar$, 
we find
\begin{eqnarray}
\label{eq:Stot}
S = Z_{f} S_{ee} + Z_{b}^{2} S_{ii} + 2 Z_{b}\sqrt{Z_{f}}S_{ei} ~.
\end{eqnarray} 
We made the assumption that the number of nuclei $n$ equals the number
of ions $n_{i}$, i.e., there are no neutral atoms in the plasma.
Eq.~(\ref{eq:Stot}) corresponds to the first two terms in
Eq.~(\ref{eq:Chihara}). In~(\ref{eq:Chihara}), however, the
contribution from the electrons screening the ion charge is separated
from $S_{ee}$. This can be done rewriting
\begin{eqnarray}
\label{eq:chi_ee_0}
\nonumber
\chi_{ee}^{RPA} 
&=& \widetilde{\chi}_{ee}^{RPA} +
\Delta\widetilde{\chi}_{ee}^{RPA}
\\
&=& \frac{\chi_{e}^{0}}%
{1-\chi_{e}^{0}\Omega_{0} V_{ee}}
+ \frac{\Omega_{0}^{2} V_{ee} V_{ii}(\chi_{e}^{0})^{2}\chi_{i}^{0}}%
{\left( 1-\chi_{e}^{0}\Omega_{0} V_{ee}-\chi_{i}^{0}\Omega_{0} V_{ii}
\right)
\left( 1-\chi_{e}^{0}\Omega_{0} V_{ee} \right)}
\nonumber\\
&=& \frac{\chi_{e}^{0}}%
{1-\chi_{e}^{0}\Omega_{0} V_{ee}}
+ \frac{(\chi_{ei}^{RPA})^{2}}{\chi_{ii}^{RPA}} ~.
\end{eqnarray}
In Eq.~(\ref{eq:chi_ee_0}) $\widetilde{\chi}_{ee}^{RPA}$ contains the
contributions from the free electrons, whereas
$\Delta\widetilde{\chi}_{ee}^{RPA}$ describes the quasi--free electrons
which screen the ions.  
From the Eqs.~(\ref{eq:chi_bb})--(\ref{eq:chi_ee_0}) we find for the
total response function
\begin{eqnarray}
\label{eq:chi_ee_tot2}
\chi_{ee}^{tot}({\bf k}, \omega) 
= \widetilde{\chi}_{ee}^{RPA}
+ \left( Z_{b} + \frac{\chi_{ei}^{RPA}}{\chi_{ii}^{RPA}}
\right)^{2} \chi_{ii}^{RPA} ~.
\end{eqnarray}

Eq.~(\ref{eq:chi_ee_tot2}) is a simple model for the total electron
response function with tightly and weakly bound electrons ``frozen''
to the ions. Since we are interested in collective Thomson scattering,
i.e. the scattering off electron fluctuations over a distance much
larger than the ion size, all atomic details are integrated
out. In RPA, Eq.~(\ref{eq:chi_ee_tot2}) is equivalent to the first two
terms in Eq.~(\ref{eq:Chihara}) with $Z_{b}=\lim_{k\to 0}f_{I}(k)$
the integrated charge of the tightly bound electrons.

Beyond the RPA, collisions are considered in the response functions
$\chi$ by utilizing the Mermin ansatz\,\cite{Mermin70} which makes use
of a relaxation time $\tau$. As outlined
in\,\cite{Selchow01,Reinholz00} the inclusion of local particle number
conservation leads to a generalization of the Mermin ansatz in terms
of a complex valued dynamical collision frequency $\tau\to
1/\nu(\omega)$. The collisional response function $\chi_{c}^{\nu,0}$
of plasma species $c$ is then given by
\begin{eqnarray}
\label{eq:MerminAnsatz2}
\chi_{c}^{\nu,0}(\mathbf{k},\omega) 
= \left( 1 - \frac{i\omega}{\nu(\omega)} \right)
\left(
\frac{\chi_{c}^{0}(\mathbf{k},z)\chi_{c}^{0}(\mathbf{k},0)}%
{\chi_{c}^{0}(\mathbf{k},z)-\frac{i\omega}{\nu(\omega)}\chi_{c}^{0}(\mathbf{k},0)}
\right)
\end{eqnarray}
and $z=\omega-{\rm Im}\,\nu(\omega)+i {\rm
  Re}\,\nu(\omega)$. Eq.~(\ref{eq:MerminAnsatz2}) is exact in the
long wavelength limit and serves as a good approximation for finite
$k$. The details of the microscopic calculation of the dynamical
collision frequency are reviewed in section\,\ref{subsec:coll_frequqncy}.
Having introduced electron--ion collisions for the OCP plasma according
to Eq.\,(\ref{eq:MerminAnsatz2}), the equations (\ref{eq:chi_ee}) --
(\ref{eq:chi_ee_tot2})
can be generalized accordingly and we obtain the total collisional
response function $\chi_{cc'}^{\nu,tot}$ by simply replacing $\chi_{c}^{0}\to
\chi_{c}^{\nu,0}$ in these equations.

\subsection{Collision Frequency}
\label{subsec:coll_frequqncy}
As outlined, e.g. in\,\cite{Selchow01,Reinholz00}, the original Ansatz
for the dielectric plasma response made by Mermin\,\cite{Mermin70},
can be generalized by calculating the dynamic collisions frequency.
In the long--wavelength limit, the dynamical collision frequency can
be expressed in terms of the inverse response function via a
generalized Drude expression. This inverse response function can be
obtained by a perturbative evaluation of the corresponding
force--force correlation function with respect to the plasma
interaction. The result for the collision frequency in Born
approximation 
can be written as (see\,\cite{Selchow01,Reinholz00} for further details)
\begin{eqnarray}
\label{eq:CollFrequ}
\nu^{\mathrm{Born}}(\omega)
= -i\mathcal{K}\frac{n_i}{n_e\omega}
\int_0^{\infty}dq\,q^6V_D^{2}(q)S_{ii}(q)
[\epsilon_{\mathrm{RPA}}(q,\omega)-\epsilon_{\mathrm{RPA}}(q,0)]
\end{eqnarray}
with
\begin{math}
\mathcal{K}=(\epsilon_0\Omega_0^{2})/(6\pi^2e^2 m_e),
\end{math}
and 
\begin{math}
S_{ii}(q)
\end{math}
the static ion--ion structure factor and
$V_{D}(q) = -Ze^{2}/(\epsilon_{0}\Omega_{0}(q^{2}+\kappa_{sc,e}^{2}))$
the statically screened Debye potential.
The electronic screening length $\kappa_{sc,e}$ is defined in
Eq.~(\ref{eq:alpha}). 
The electron RPA dielectric
function $\epsilon^{RPA}$ in Eq.~(\ref{eq:CollFrequ}) is related to
the Lindhard response function, Eq.~(\ref{eq:Lindhard}), according to 
\begin{math}
\epsilon_{\mathrm{RPA}}^{-1}=1+\chi_{e}^{0}e^{2}/(\epsilon_{0}k^{2}).
\end{math} 
It should be pointed out that
the electron screening is accounted for in the Debye potential,
whereas ion screening is treated in the ion--ion structure factor
according to 
\begin{eqnarray}
\label{eq:SiiStat}
S_{ii}(q) = \frac{q^{2}}{q^{2}+\kappa_{D,i}^{2}} ~,
\end{eqnarray}
with the ion Debye screening length
\begin{math}
\kappa_{D,i}^2 =
Z_{f}^{2} n_{i}e^{2}/(\epsilon_{0}k_{B}T_{i}).
\end{math}
Improvements of $S_{ii}$ which account for ion--ion correlations and
screening effects are given, e.g., using pseudo--potentials in
hypernetted chain calculations
(HNC)\,\cite{Seuferling89,Arkhipov98,Gregori06,Schwarz06}. These
improvements go beyond the scope of this paper, where we are
interested in the electron plasma component only, which is determined by the
high--frequency electron collective mode (plasmons).  For an improved
plasma diagnostics which measures the electron and ion component of the
plasma separately from the scattering spectrum, an advanced theory to
describe the ion--ion correlations, like HNC, is needed. This is an
ultimate requirement if probing nonequilibrium plasma states, e.g.,
in a two--temperature plasma with $T_{e}\ne T_{i}$.

\subsection{Dynamical Structure Factor and Plasma Properties}
\label{subsec:SProperties}
The dynamical structure factor shows distinct features, which are
fingerprints of the plasma properties. Therefore, collective
Thomson scattering can be applied as a diagnostic tool as described.
The relation of the scattering spectrum (or dynamical structure
factor) to the plasma properties is explained in the following.

\begin{figure}[ht]
\begin{center}
\begin{minipage}{0.8\textwidth}
\includegraphics[angle=270,width=\textwidth]{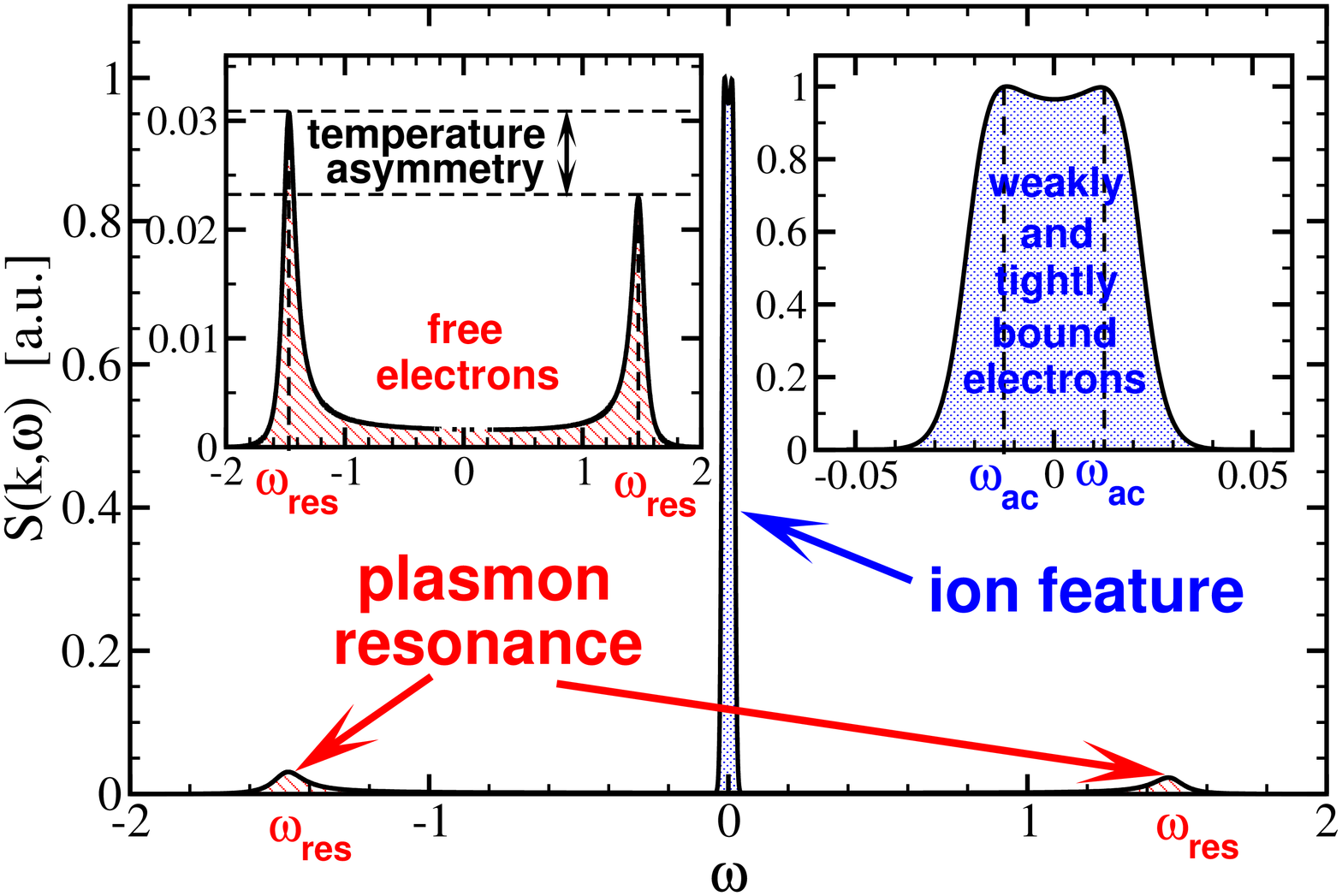}
\end{minipage}
\\
\caption{\label{fig:DynStructFactorSchematic}Schematic view of the
  dynamical structure factor $S(\mathbf{k},\omega)$ as a function of
  the frequency shift $\omega$ in the collective scattering regime
  (color online). The high resonance collective mode (plasmons) is
  shown as a red hatched region. At low frequency shifts, the ion
  feature due to weakly and tightly bound electrons is shown as the
  blue dotted region. The plasmon and ion acoustic resonance
  frequency, $\omega_{res}$ and $\omega_{ac}$, respectively, are
  shown. The insets on the upper left and right show magnifications of
  the plasmon resonance and the ion feature respectively.}
\end{center}
\end{figure}

In Fig.~\ref{fig:DynStructFactorSchematic} we show a schematic picture
of the dynamical structure factor in the collective scattering regime
with its characteristic features, the ion feature and the plasmon
resonance.
The asymmetry in the height of the dynamical structure factor at
$\pm\omega$ can be explained as follows.  The scattering of a photon
with initial momentum $\mathbf{k}_{i}$ and frequency $\omega_{i}$ into
final momentum $\mathbf{k}_{f}$ and frequency $\omega_{f}$ must be
proportional to the probability of the final photon state. Assuming
that the system is in thermal equilibrium at a temperature $T$, not
disturbed by the laser irradiation, the dynamical structure factor has
to be proportional to the Boltzmann factor according to
\begin{eqnarray}
\label{eq:Scatt_12}
S(\mathbf{k}_{i}\to\mathbf{k}_{f},\omega_{i}\to\omega_{f}) \propto
\exp\left\{ -\frac{\hbar\omega_{f}}{k_{B}T} \right\} ~.
\end{eqnarray} 
Similar, for the reverse scattering we have
\begin{eqnarray}
\label{eq:Scatt_21}
S(\mathbf{k}_{f}\to\mathbf{k}_{i},\omega_{f}\to\omega_{i}) \propto
\exp\left\{ -\frac{\hbar\omega_{i}}{k_{B}T} \right\} ~.
\end{eqnarray} 
For a homogeneous and stationary system, the scattering process is
described by $\mathbf{k}\equiv \mathbf{k}_{f}-\mathbf{k}_{i}$ and
$\omega\equiv\omega_{f}-\omega_{i}$, the momentum and energy
transfer, only. From the general property\,\cite{Sitenko95}
$\chi(\mathbf{k},\omega)=\chi^{\ast}(-\mathbf{k},-\omega)$%
\footnote{This is a general requirement due to the reality of the
  induced current $\mathbf{j}$ and the vector potential $\mathbf{A}$
  acting as the external perturbation.},
we find via the equilibrium FDT (\ref{eq:FDT}) and the relations
(\ref{eq:Scatt_12}), (\ref{eq:Scatt_21})
\begin{eqnarray}
\label{eq:DetBalance}
\frac{S(-\mathbf{k},-\omega)}{S(\mathbf{k},\omega)} 
= \exp\left( -\frac{\hbar\omega}{k_{B}T} \right) ~.
\end{eqnarray}
Eq.~(\ref{eq:DetBalance}) is known as detailed balance relation, and
it is a general property following from first principles. As seen, the
ratio defined in (\ref{eq:DetBalance}) at any particular energy
transfer $\hbar\omega$ depends only on the equilibrium temperature of
the system. This makes the detailed balance relation a highly suitable
tool to measure the temperature of the plasma, independent
of any model assumptions other than thermodynamic equilibrium. In
Fig.~\ref{fig:DynStructFactorSchematic} the asymmetry in the dynamical
structure factor, denoted as temperature asymmetry, is due to the
detailed balance relation.

Scattering from electrons which are weakly or tightly bound to the
ions, see discussion in connection with Eq.~(\ref{eq:Chihara}), leads
to the low frequency resonance, the ion feature. As shown in
Fig.~\ref{fig:DynStructFactorSchematic}, due to the large ion mass,
the ion feature shows a very narrow spectral width.  Currently, there
is no x--ray or VUV radiation available that would allow to spectrally
resolve the ion feature. In the future this could be possible with FEL
seeding\,\cite{Feldhaus97} and in recently proposed FELs based on a
energy recovery linac scheme with very narrow XUV
bandwidth\,\cite{4gls}.

The high--frequency mode, known as plasmon, is due to the collective
scattering from free electrons. In
Fig.~\ref{fig:DynStructFactorSchematic}, a red ($\omega<0$) and blue
($\omega>0$) shifted plasmon feature appear due to the creation and
annihilation of a plasmon, respectively.  Neglecting the plasmon
collisional and Landau damping, the plasmon resonance $\omega_{res}$
is approximately given by the plasmon dispersion relation.  For a
classical collisionless plasma an expression for the plasmon
dispersion has been given by Gross and Bohm\,\cite{Gross49}, valid in
the long wavelength limit $\hbar^{}k^{2}/(2m_{e}\omega)\ll 1$
\begin{eqnarray}
\label{eq:GrossBohm}
\omega_{res}^{2}\approx \omega_{pe}^{2} +
\frac{3k_{B}T_{e}}{m_{e}}k^{2} ~,
\end{eqnarray}
with the density dependent plasma frequency $\omega_{pe}^{}$.  The
dispersion relation is calculated from the condition ${\rm
  Re}\,\epsilon(\mathbf{k},\omega)=0$. Eq.~(\ref{eq:GrossBohm})
follows for a classical Maxwell Boltzmann plasma and the dielectric
function $\epsilon$ expanded to order $\mathcal{O}(k^2)$.  An
inspection of Eq.~(\ref{eq:GrossBohm}) reveals immediately that the
plasmon resonance position is mainly determined by the plasma
frequency and, therefore, by the electron density. The second term in
the dispersion relation contains a temperature dependence. Having
determined the temperature via the detailed balance relation as
explained, the measurement of the plasmon position $\omega_{res}$
provides the free electron density in the plasma.

An improvement of the classical dispersion relation
(\ref{eq:GrossBohm}) including quantum diffraction is obtained if the
electron Lindhard expression for the dielectric function is solved for
${\rm Re}\,\epsilon(\mathbf{k},\omega)=0$. The second moments of the
Fermi function are given by Fermi integrals, which can be expressed
using an approximate solution\,\cite{Zimmermann98} to the lowest order
in $y=n_{e}\Lambda_{e}^{3}/g_{e}$. Here we have defined the electron
thermal wavelength $\Lambda_{e}=h/\sqrt{2\pi m_{e}k_{B}T_{e}}$ and
$g_{e}$ the electron spin degeneracy factor. With these approximations
an improved dispersion relation (IDR) can be obtained
\begin{eqnarray}
\label{eq:IDR}
\omega_{res}^{2}(k^{2})\approx \omega_{pe}^{2} +
\frac{3k_{B}T_{e}}{m_{e}}k^{2} 
\left( 1 + 0.088n_{e}\Lambda_{e}^{3} + \ldots \right)
+\left( \frac{\hbar k^{2}}{2m_{e}} \right)^{2} ~,
\end{eqnarray}
and the ellipsis denoting higher order terms in $y$. This expansion is
valid for $y<5.5$\,\cite{Zimmermann98}. At the order
$\mathcal{O}(k^{4})$ only the leading term is kept. Other terms of the
order $\mathcal{O}(k^{4})$\,\cite{Kraeft86} are suppressed by higher
moments of the distribution function. It is a simple observation from
Eq.~(\ref{eq:IDR}) that in the limit $\hbar\to 0$ and
$n_{e}\Lambda_{e}^{3}\to 0$ the Gross--Bohm dispersion relation,
Eq.~(\ref{eq:GrossBohm}) is recovered. Further, it must be pointed out
that both dispersion relations assume ${\rm
  Im}\,\epsilon(\mathbf{k},\omega)=0$ and, consequently, neglect the
effect of collisional damping.

\subsection{Continuum Radiation}
\label{subsec:ResContRad}
Independent information on the plasma density and temperature can
be obtained from the continuum (bremsstrahlung) emission of the
plasma.  For demonstration, the emission for an optically thin
hydrogen plasma ($Z_{f}=1$) in thermal equilibrium at different
temperatures is given by the classical Kramers
formula~\cite{Kramers23,Rybi75}
\begin{eqnarray}
\label{eq:Kramers}
\frac{\mathrm{d}j(\lambda)}{\mathrm{d}V\mathrm{d}\Omega}
=\frac{\mathrm{d}P_{b}(\lambda)}{\mathrm{d}V\mathrm{d}\lambda\mathrm{d}\Omega}
=\mathcal{C}_{b}\frac{Z g_{ff}n_e^2e^{6}}{3m_ec^2\lambda^2}
\left( \frac{2\pi}{3 k_\mathrm{B}T m_e }
\right)^{1/2}
\exp\left( -\frac{2\pi\hbar c}{\lambda k_\mathrm{B}T} \right)~,
\end{eqnarray}
with 
\begin{math}
\mathcal{C}_{b}=2^{5}\pi/(4\pi \epsilon_{0})^3
\end{math}
and $g_{ff}$, the Gaunt factor\,\cite{Gaunt30}, accounting for quantum
and medium corrections. A quantum mechanical calculation of the Gaunt
factor\,\cite{Fortmann06} showed that $g_{ff}=1$ serves as a good
approximation for the spectrum.

\section{Results}
\label{sec:Results}
\subsection{Collision Frequency}
\label{subsec:ResCollFrequency}
\begin{figure}[ht]
\begin{center}
\begin{minipage}{0.8\textwidth}
\includegraphics[angle=270,width=\textwidth]{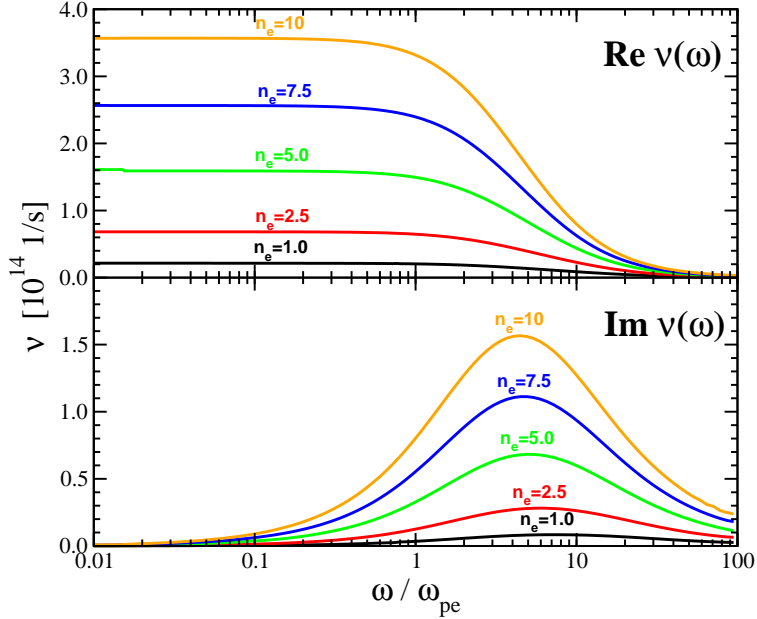}
\end{minipage}
\\
\caption{\label{fig:CollFrequBorn}Real and imaginary part of the
  electron ion dynamical collision frequency in hydrogen as a function
  of the energy transfer (upper and lower panel respectively). The
  collision frequency is calculated in Born approximation,
  Eq.~(\ref{eq:CollFrequ}), with respect to a statically screened
  Debye potential. The electron and ion temperature is $12\,{\rm eV}$
  and the free electron density $n_{e}$ is given in units of
  $10^{21}\,{\rm cm}^{-3}$.}
\end{center}
\end{figure}

Using the results of the previous section we can calculate the
dynamical structure factor $S(\mathbf{k},\omega)$ via the FDT,
Eq.~(\ref{eq:FDT}), for a collisional hydrogen plasma.  The result for
the collision frequency calculated in Born approximation according
to Eq.~(\ref{eq:CollFrequ}) is shown in Fig.~\ref{fig:CollFrequBorn}.
It is seen that collsions become more important at higher densities
in this domain. This is due to an increasing coupling parameter.  
At frequencies larger than the plasma frequencies $\omega_{pe}$, the
collisions become less effective and the real and imaginary part of
the collision frequency goes to zero.

\subsection{Dynamical Structure Factor}
\label{subsec:ResDynStructFact}
\begin{figure}[ht]
\begin{center}
\begin{minipage}{0.8\textwidth}
\includegraphics[angle=270,width=\textwidth]{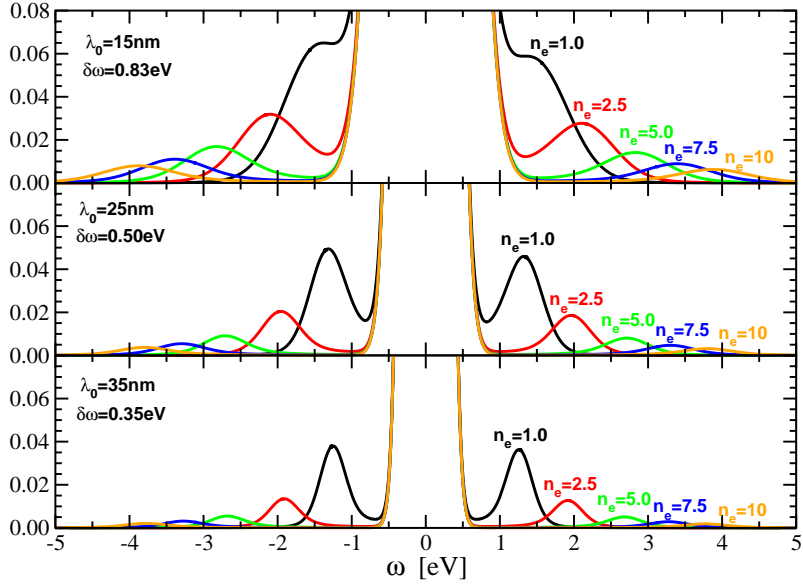}
\end{minipage}
\\
\caption{\label{fig:DynStructFactor}Hydrogen dynamical structure factor
  $S(k,\omega)$ in units of $1/{\rm eV}$ calculated from
  Eqs.~(\ref{eq:Stot}), (\ref{eq:MerminAnsatz2}) and
  (\ref{eq:CollFrequ}) including collisions in Born approximation.
  Each panel shows $S$ for different electron densities $n_{e}$ in
  units of $10^{21}\,{\rm cm}^{-3}$, at equilibrium conditions
  $T_{e}=T_{i}=12\,{\rm eV}$, ionization degree $Z_{f}=1$ and a scattering
  angle of $\theta=90^{\circ}$. For comparison three VUV probe
  wavelength ($15$, $25$ and $35\,{\rm nm}$) are shown. The finite
  VUV--bandwidth as well as the spectrometer resolution is accounted
  for by applying Gaussian convolution with $\delta\omega /
  \omega_{0}=0.01$ FWHM.}
\end{center}
\end{figure}

From the collision frequencies we have calculated the dynamical
structure factor using Eqs.~(\ref{eq:chi_ee_tot2}) with the
replacement $\chi_{c}^{0}\to\chi_{c}^{\nu,0}$ for an equilibrium
plasma at $T_{e}=T_{i}=12\,{\rm eV}$. The results are plotted in
Fig.~\ref{fig:DynStructFactor} as a function of the frequency shift
$\omega$. Three different FEL probe wavelengths $15\,{\rm nm}$,
$25\,{\rm nm}$ and $35\,{\rm nm}$ are shown in the upper, middle and
lower panel, respectively. The corresponding momentum transfer
$\mathbf{k}$ is calculated from Eq.~(\ref{eq:MomTransfer}) for a
scattering angle of $90^\circ$.
The structure factor is calculated for the same electron densities as
in Fig.~\ref{fig:CollFrequBorn}. In order to account for the FEL
bandwidth and the final spectrometer resolution, we convoluted the
structure factor with a Gaussian profile of $\delta\omega/\omega=0.01$
full width half maximum (FWHM). In Fig.~\ref{fig:DynStructFactor}
elastic scattering of the bound electrons at $\omega=0$ contributes to
the ion feature. Its spectral width is given by the bandwidth and
spectrometer resolution. Therefore, the ion feature serves as a
measure of the FEL bandwidth. Beside the ion feature we observe the
blue and red shifted plasmon resonance.  Both resonances are different
in height, attributed to the detailed balance relation,
Eq.~(\ref{eq:DetBalance}). This asymmetry increases with decreasing
temperature.  In accordance with the Gross--Bohm and the improved
dispersion relation, Eqs.~(\ref{eq:GrossBohm}) and (\ref{eq:IDR}), one
can observe that the plasmon resonance is shifted to larger resonance
frequencies with increasing densities due to an increase of the plasma
frequency. Further, in agreement with the plasmon dispersion relation,
a larger probe wavelength, leading to a smaller momentum transfer,
Eq.~(\ref{eq:MomTransfer}), yields a smaller plasmon resonance
frequency.  A calculation of the dynamical structure factor in the
density range $10^{21}\,{\rm cm}^{-3}<n_{e}<10^{22}\,{\rm cm}^{-3}$
and a subsequent calculation of the electron density $n_{e}^{GB}$ from
the plasmon resonance position via the Gross--Bohm dispersion relation
yields a deviation of $(n_{e}^{BG}-n_{e})/n_{e}<0.15$. This deviation
is attributed to collisional and quantum effects as well as details of
the electron distribution functions, not accounted for in the
Gross--Bohm dispersion relation. It should be pointed out that this
deviation increases significantly to more than $0.4$ if x--ray
scattering is considered at solid densities as, for instance,
scattering conditions in\,\cite{Glenzer06}.

\subsection{Continuum Radiation}
\label{subsec:ResContRad2}
\begin{figure}[ht]
\begin{center}
\begin{minipage}{0.8\textwidth}
\includegraphics[angle=270,width=\textwidth]{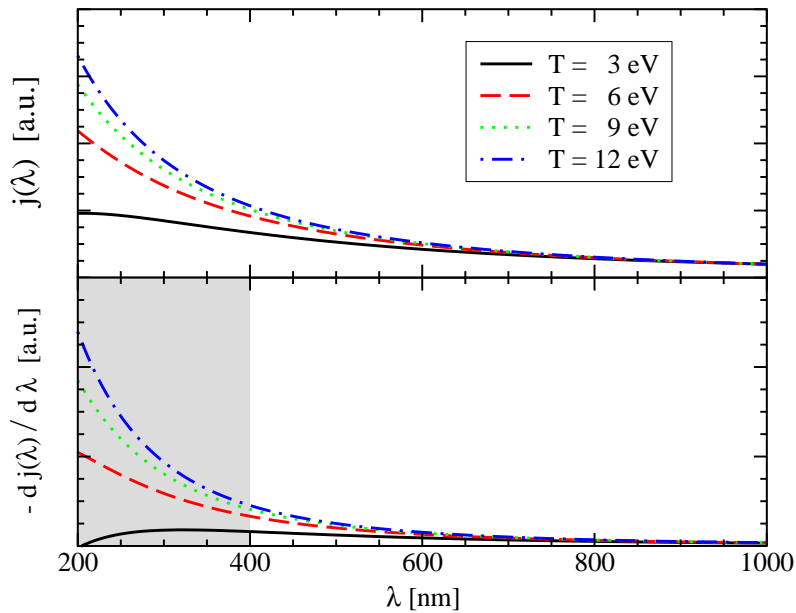}
\end{minipage}
\\
\caption{\label{fig:Brems}Bremsstrahlung and its negative derivative for a
  Gaunt factor $g_{ff}=1$ for different temperatures at constant
  density. Both curves are normalized at $\lambda=1000\,{\rm nm}$ and
  plotted in arbitrary units.The gray region shows the wavelengths
  where the spectrum is sensitive to the temperature.}
\end{center}
\end{figure}

In order to estimate the sensitivity of the bremsstrahlung emission
with respect to the temperature
we plot Eq.~(\ref{eq:Kramers}) and its derivative in
Fig.~\ref{fig:Brems} for a range of different temperatures. Both
panels are normalized at $\lambda=1000\,{\rm nm}$ and are given in
arbitrary units.
Since we assume no absolute calibration of the spectrum, only the
derivative of the spectrum can provide information (lower panel in
Fig.\,\ref{fig:Brems}). The derivative of the spectrum for different
temperatures shows deviations at small wavelengths and, consequently,
there is a temperature sensitivity of the spectrum present. At small
wavelengths the plasma is optically thin and Eq.~(\ref{eq:Kramers})
is applicable. Therefore, measuring the continuum spectrum and fitting
the spectrum for a given electron density (measured by collective
Thomson scattering) yields an independent estimate of the plasma
temperature.

\section{Thomson Scattering Experiment at FLASH}
\label{sec:Experiment}
To demonstrate the scattering from the collective electron mode of an
equilibrium, near--solid density plasma at moderate temperatures
$T=1-15\,{\rm eV}$, we propose a proof--of--principle experiment at the
VUV free--electron laser FLASH at DESY, Hamburg. 
The aim of the experiment consists of (i) creating a plasma from a low
$Z$ target at near--solid density by an optical heating laser. After a
relaxation time in a second step (ii) the plasma is probed by the
VUV--FEL and the scattered spectrum is measured by a high resolution
transmission grating spectrometer.  This novel pump--probe experiment
allows to determine basic plasma properties from the distinct features
of the scattering spectrum as outlined in the previous sections.

\subsection{Experimental Requirements}
\label{subsec:FLASH_requirements}
Based on our results described in section~\ref{sec:Results}, we have
analyzed the realization of such an experiment and list some of
its most important features: 
\begin{enumerate}
\item In order to obtain a strong plasmon scattering signal resulting
  from free electrons and a preferably weak signal from the bound
  electrons, a low $Z$ target material is used.
\item As the target we employ a cryogenic hydrogen beam which,
  upon full ionization, provides a free electron density of
  $n_{e}=(2.2-2.4)\times 10^{22}\,{\rm cm}^{-3}$. After a relaxation
  time of about $1\,{\rm ps}$ we expect a plasma density of
  about $n_{e}=10^{21}-10^{22}\,{\rm cm}^{-3}$.
\item Making use of a scattering geometry $\varphi=90^{\circ}$,
  $\theta=90^{\circ}$ as shown in Fig.~\ref{fig:PolarizationScheme}
  and \ref{fig:TargetChamber}, the linear polarization of the FEL
  pulse does not decrease the scattering spectrum (see
  Fig.~\ref{fig:PolarizationDependence}).  In a temperature range
  $T_{e}=1-15\,{\rm eV}$ a collective scattering spectrum is expected,
  as exemplarily shown in Fig.~\ref{fig:DynStructFactor} for a
  temperature of $T_{e}=12\,{\rm eV}$ and various electron densities.
\item Given that the electron density is in the range
  $n_{e}=10^{21}-10^{22}\,{\rm cm}^{-3}$, FEL radiation at a
  wavelength of $\lambda_{0}=25\,{\rm nm}$ ($\approx50\,{\rm eV}$) is
  chosen in order to fully separate the plasmons from the ion feature.
  Due to the bandwidth characteristics of the FEL, this cannot be
  achieved at a smaller wavelength at, e.g.  $\lambda_{0}=15\,{\rm
    nm}$ (see Fig.~\ref{fig:DynStructFactor}).
\item The application of the detailed balance relation,
  Eq.~(\ref{eq:DetBalance}), for the temperature measurement restricts
  the plasma temperature to $T_{e}\lesssim 15\,{\rm eV}$. At larger
  temperatures the asymmetry in the scattering spectrum becomes less
  than $10\%$ and cannot be resolved experimentally.
\item The number of photons scattered from the plasmons has to be
  sufficiently high. For a FLASH pulse of
  energy $E_{FLASH}$ and wavelength $\lambda_{0}$ focused onto the
  target, one obtains the total number of photons $N_{ph}^{tot}$ by
\begin{eqnarray}
\label{eq:Ntot}
N_{ph}^{tot} 
= 5.03\times 10^{9}\,E_{FLASH}[\mu{\rm J}] \lambda_{0}[{\rm nm}] ~.
\end{eqnarray}
The scattered fraction for a plasma length $L$ and the scattering
cross section
$\sigma=\sigma_{T}/(1+\alpha^{2})\approx\sigma_{T}/\alpha^{2}$ (note:
$\alpha>3$) with $\sigma_{T}=6.65\times10^{-25}\,{\rm cm}^2$ the total
Thomson cross section, is given by $\sigma n_{e}L$. Neglecting the
density correction in the momentum transfer,
Eq.~(\ref{eq:MomTransfer}), we find with Eq.~(\ref{eq:alpha}) for the
scattering parameter $\alpha$
\begin{eqnarray}
\label{eq:alpha2}
  \alpha^{2} = 0.1146\; 
  \frac{\lambda_{0}^{2}[{\rm nm}^{2}]n_{e}[10^{21}{\rm cm}^{-3}]}%
  {T_{e}[{\rm eV}]\sin^{2}\theta/2} ~.
\end{eqnarray}
Therefore, the number of scattered photons $N_{ph}^{sc}$ into the
acceptance solid angle of the detector $\Delta\Omega=6\times
10^{-4}\,{\rm sr}$ is given by
\begin{eqnarray}
\label{eq:NphotScattered}
N_{ph}^{sc} \approx 1.753\;\frac{E_{FLASH}[\mu{\rm J}] L[\mu{\rm
    m}]T_{e}[{\rm eV}]}{\lambda_{0}[{\rm nm}]\sin^{2}\theta/2} ~.
\end{eqnarray}
For a plasma length $L=40\,\mu{\rm m}$, a wavelength
$\lambda_{0}=25\,{\rm nm}$, a pulse energy $E_{FLASH}=30\,\mu{\rm J}$,
and a scattering angle of $\theta=90^{\circ}$ we find the number of
photons collected with the spectrometer $N_{ph}^{sc} \approx
170\,T_{e}{[{\rm eV}]}$.
\end{enumerate}

\subsection{Experimental Setup}
\label{subsec:ExperimentalSetup}
\begin{figure}[ht]
\begin{center}
\includegraphics[width=0.7\textwidth,angle=0]{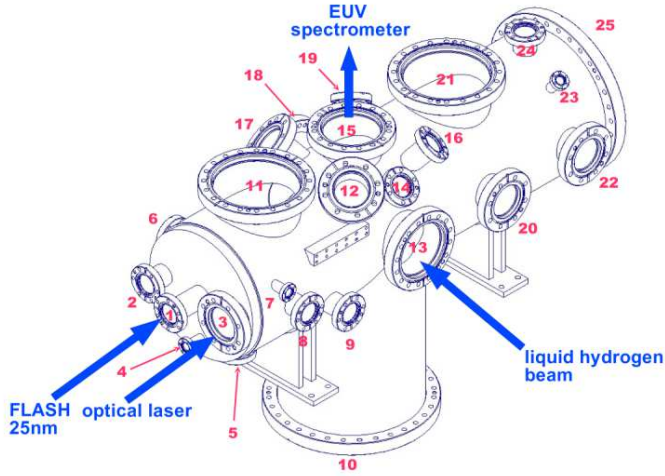}
\caption{\label{fig:TargetChamber}Vacuum chamber with the main connections.}
\end{center}
\end{figure}

\begin{figure}[ht]
\begin{center}
\includegraphics[width=0.7\textwidth,angle=0]{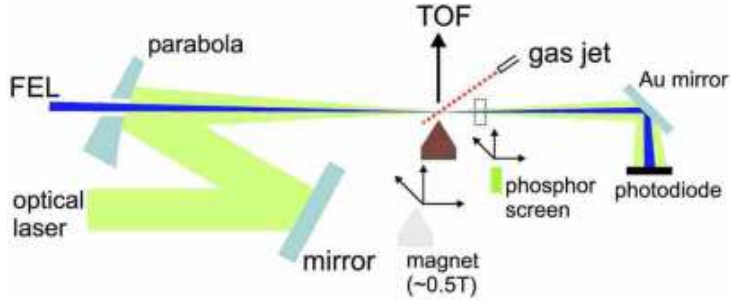}
\caption{\label{fig:LaserSynchronization}Adjustment of the space and
  time overlap of the FEL and optical laser.}
\end{center}
\end{figure}

The vacuum chamber with its main connections being used for the
Thomson scattering experiment at DESY and the alignment of the laser
beams are shown in Fig.~\ref{fig:TargetChamber} and
Fig.~\ref{fig:LaserSynchronization}, respectively.
The plasma is generated by $100\,{\rm fs}$ pulses with up to $10\,{\rm
  mJ}$ energy at $800\,{\rm nm}$ which are focused to a spot size of
$50\,\mu{\rm m}$ in the interaction region where the cryogenic
hydrogen beam crosses perpendicularly to the laser beams.  For probing
the plasma, the FEL beam enters the interaction region through a hole
in the parabola mirror, having a beam waist diameter of $20\,\mu{\rm
  m}$ in the interaction region. Important for the success of the
experiment is a good control of the pump--probe--delay between the
optical pump laser and the FEL. A parallel alignment of the beams, as
shown in Fig.~\ref{fig:LaserSynchronization}, is chosen in order to
simplify the coarse adjustment of the temporal overlap using a fast
photodiode. For the fine adjustment, the photoelectron (PE) sideband
generation technique is applied, which has been successfully
demonstrated at FLASH\,\cite{Meyer05} at a helium gas jet. We will use
the same laser beam alignment (note:
Fig.~\ref{fig:LaserSynchronization} is taken from\,\cite{Meyer05}) and
a similar PE sideband generation setup. In contrast to\,\cite{Meyer05}
we apply the technique to the hydrogen gas jet at $100\,{K}$ and
record the PE spectrum by a field free electron time--of--flight (TOF)
spectrometer.  Once the temporal overlap between the optical laser and
the FEL has been determined, the the pump--probe delay can be adjusted
up to several nanoseconds by a motorized translation stage. The
stability of the laser pulse timing is monitored by a streak camera
and can be measured on a shot to shot basis by electro--optical
sampling\,\cite{Berden04} with an accuracy of $200\,{\rm fs}$. The
repetition rate of the experiment is $5\,{\rm Hz}$, determined by the
repetition rate of the FEL.

The hydrogen beam runs perpendicular to the optical laser and the FEL
through their joint focus.  A cryogenic source similar to the one used
here has been characterized for helium at beam diameters of
$30-100\,\mu{\rm m}$\,\cite{Slipchenko02}.  Preparatory tests aiming
at a continuous hydrogen beam have been performed at the University of
Rostock.  Depending on the pressure and temperature conditions, the
source provides: a spray of hydrogen droplets, a well collimated beam
of droplets or a slow moving filament of solid hydrogen.  For the
experiment we will make use of the collimated beam consisting of
droplets $40\,\mu{\rm m}$ in diameter. It operates stable at a
temperature of $T\approx 15\,{\rm K}$ and a pressure of $p\approx
15\,{\rm bar}$ for a nozzle diameter of $d_{nozzle}=20\,\mu{\rm m}$.
Similar studies on cryogenic argon beams\,\cite{Grams05} and
piezo-driven hydrogen droplets\,\cite{Trostell95} have been reported.

\begin{figure}[ht]
\begin{center}
\includegraphics[width=0.7\textwidth,angle=0]{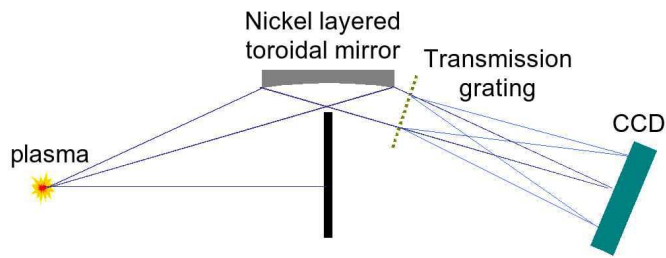}
\caption{\label{fig:VUVSectrometer}Schematic view of the EUV spectrometer.}
\end{center}
\end{figure}

As shown in Fig.~\ref{fig:TargetChamber}, the scattered photons are
measured by a high resolution transmission grating EUV spectrometer,
mounted perpendicular on top of the interaction region at a scattering
angle of $\theta=90^{\circ}$. The spectrometer layout is
schematically shown in Fig.~\ref{fig:VUVSectrometer},
details are reported in\,\cite{Jasny94}.  A large area transmission
grating with a line density of $1000\,{\rm lines/mm}$ will be
used. The spectral range is $\lambda=0.5-50\,{\rm nm}$ and the
expected spectrum is optimally distributed on the CCD surface (size:
$27\times13.5\,{\rm mm}^{2}$). The spectral resolution of the
spectrometer is mainly limited by the dimension of the plasma and the
pixel size of the soft x--ray CCD ($13.5\times13.5\,\mu{\rm m}^{2}$).
At the wavelength $\lambda=25\,{\rm nm}$ we expect a resolution of
$\Delta\lambda/\lambda\approx8\times10^{-3}$. The acceptance solid
angle is $\Delta\Omega=6\times10^{-4}\,{\rm sr}$.

As additional diagnostics an optical spectrometer for the detection of
continuum radiation and a CCD--based microscope pointing to the focus
will be employed. The microscope is used for characterization of the
hydrogen beam in terms of droplet number, size and speed and for
initial alignment.

\section{Summary}
\label{sec:Summary}
We have shown that x--ray Thomson scattering can be applied as a
diagnostic tool to elaborate correlated plasmas in the WDM regime. In
particular, we propose a proof--of--principle experiment at the VUV
FLASH facility at DESY to measure equilibrium properties of a
near--solid density hydrogen plasma. High--brilliant coherent VUV FEL
radiation at a wavelength of $\lambda_{0}=25\,{\rm nm}$ will be used
to scatter off the dense plasma. The collective Thomson scattering
spectrum will be measured using a high resolution, high sensitivity
transmission grating EUV-spectrometer. As our theoretical calculations
show, the distinctive plasmon features expected in the spectrum, will
allow to measure the electron density and temperature. The plasma
regime considered in this paper lies in the range of
$n_{e}=10^{21}-10^{22}\,{\rm cm}^{-3}$ for the free electron density
and $T_{e}=1-15\,{\rm eV}$ for the temperature. A plasma under these
conditions is generated by laser irradiation of a cryogenic hydrogen
droplet beam. A time delay between the pump and probe laser of the
order of the relaxation time ($\approx 1\,{\rm ps}$) will allow to
probe the plasma at equilibrium condition.

The electron temperature will be determined by the asymmetry between
the red and blue shifted part of the spectrum.  This property is
directly related to the detailed balance relation which is based on
first principles. Therefore, the method provides a reliable measure of
the equilibrium temperature.  The position of the plasmon resonance is
determined by the density. A calculation of the dynamical structure
factor for different electron densities allows to identify the plasmon
resonance position by the maximum of the plasmon feature. This method
provides a reliable electron density measurement if collisional and
Landau damping as well as quantum statistical effects are accounted
for in the calculations. Estimations for the density can be obtained
based on the plasmon dispersion relation with a systematic error of
less than $15\%$ in the considered plasma regime.  Limiting factors of
the proposed measurement have been estimated. The spectral resolution,
resulting from the finite FEL bandwidth and EUV--spectrometer
resolution, is sufficiently large to spectrally separate the plasmon
from the Rayleigh peak. The number of scattered photons off the
plasmons has been estimated. More information will be obtained from
additional diagnostics.  For instance, the measurement of the
continuum radiation provides independent information on the plasma
density and temperature.

The proposed plasma diagnostic method can be developed into a standard
tool for WDM research. Due to the unique features of x-ray and VUV FEL
radiation concerning brilliance and coherence, they are favorable
sources for Thomson scattering diagnostics which has the potential to
be extended to nonequilibrium plasmas in the future. This will need a
time--resolved scattering experiment which will reveal the plasma
dynamics in WDM and experiments are currently proposed, e.g. at
FLASH\,\cite{Hoell06}.  The experiment proposed in this paper serves
as a precursor to a systematic elaboration of the widely unexplored
regime of WDM.

\section{Acknowledgments}
\label{sec:Acknowledgments}
This work was supported by the virtual institute VH-VI-104 of the
Helmholtz association and the Sonderforschungsbereich SFB 652. The
work of SHG was performed under the auspices of the U.S. Department of
Energy by the University of California Lawrence Livermore National
Laboratory under contract number No.  W-7405-ENG-48 and the
Alexander--von--Humboldt foundation. The work of GG was supported by
the Council for the Central Laboratory of the Research Councils (UK).


\begin{thebibliography}{00}
\bibitem{Lee03}
  R.W.~Lee, et al., 
  J. Opt. Soc. Am. B {\bf 20}, 770 (2003).
\bibitem{Riley00}
  D.~Riley, et al., 
  Phys.  Rev.  Lett. {\bf 84}, 1704 (2000).
\bibitem{Landen01}
  O.L.~Landen, et al., 
  J. Quant. Spectrosc. Radiat. Transfer {\bf 71}, 465 (2001).
\bibitem{Gregori03}
  G.~Gregori, et al., 
  Phys.  Rev.  E {\bf 67}, 026412 (2003).
\bibitem{Hoell04}
  A.~H\"oll, et al., 
  Eur. Phys. J. D {\bf 29}, 159 (2004).
\bibitem{Redmer05}
  R.~Redmer, et al., 
  IEEE Trans. Plasma Science {\bf 33} , 77 (2005).
\bibitem{Landen01_2}
  O.L.~Landen, et al.,
  Rev. Sci. Inst. {\bf 72}, 627 (2001).
\bibitem{Glenzer03} 
  S.H.~Glenzer, et al., 
  Phys. Rev. Lett. {\bf 90}, 175002 (2003).
\bibitem{Glenzer06} 
  S.H.~Glenzer, et al.,
  Phys. Rev. Lett. (2006), submitted.
\bibitem{Ayvazyan06} 
  V.~Ayvazyan, et al.,
  Eur. Phys. J. D. {\bf 37}, 297 (2006).
\bibitem{Stojanovic06}
  N.~Stojanovic, et al.,
  Appl. Phys. Lett., accepted for publication (2006).
\bibitem{Chapman06}
  H.N.~Chapman et al.,
  Nature Physics Letters, Nov. (2006).
\bibitem{Saldin80} 
  A.~Kondratenko, E.~Saldin,
  Part. Accel. {\bf 10}, 207 (1980).
\bibitem{Bonifacio84} 
  R.~Bonifacio, C.~Pellegrini,
  Opt. Commun. {\bf 50}, 373 (1984).
\bibitem{Hughes75}
  T.P.~Hughes,
  {\it Plasma and Laser Light},
  John Wiley \& Sons, NY. (1975).
\bibitem{Sheffield75}
  J.~Sheffield,
  {\it Plasma Scattering of Electromagnetic Radiation},
  Academic Press, N.Y. (1975).
\bibitem{Bekefi66}
  G.~Bekefi, 
  {\it Radiation Processes in Plasmas}, 
  John Wiley, New York (1966).
\bibitem{Klimontovich82}
  Yu.L.~Klimontovich,
  {\it Kinetic Theory of Nonideal Gases and Nonideal Plasmas},
  Pergamon, Oxford (1982).
\bibitem{Gregori04}
  G.~Gregori, et al., 
  J. Phys. A {\bf 36} , 5971 (2003).
\bibitem{Ichimaru85}
  S.~Ichimaru, et al., 
  Phys. Rev. A{\bf 32} , 1768 (1985).
\bibitem{Ichimaru94}
  S.~Ichimaru,
  {\it Statistical Plasma Physics},
  vol. II: Condensed Plasmas, Addison--Wesley (1994).
\bibitem{Roepke98}
  G.~R\"opke,
  Phys. Rev. E {\bf 57}, 4673 (1998).
\bibitem{Roepke98_2}
  G.~R\"opke, A.~Wierling,
  Phys. Rev. E {\bf 57}, 7075 (1998).
\bibitem{Wierling01}
  A.~Wierling, et al., 
  Physics of Plasmas {\bf 8}, 3810 (2001).
\bibitem{Reinholz05}
  H.~Reinholz, 
  Ann. Phys. (Fr) {\bf 30}, 1 (2005).
\bibitem{Dharma-wardana06}
  M.W.C.~Darma-Wardana, 
  Phys. Rev. E {\bf 73}, 036401 (2006).
\bibitem{Ichimaru88}
  S. Ichimaru, 
  {\it Plasma Physics: An Introduction to Statistical Physics of
    Charged Particles},
  Addison--Wesley (1988).
\bibitem{Kubo66}
  R.~Kubo,
  Rep. Prog. Phys. {\bf 29}, 255 (1966).
\bibitem{Selchow01}
  A.~Selchow, et al.,
  Phys. Rev. E {\bf 64}, 056410 (2001).
\bibitem{Chihara87}
  J.~Chihara,
  J. Phys. F {\bf 17}, 295 (1987).
\bibitem{Chihara00}
  J.~Chihara,
  J. Phys.: Cond. Matter {\bf 12}, 231 (2000).
\bibitem{Roepke79} 
  G.~R\"opke, R.~Der, 
  Phys. Stat. Sol. B {\bf 92}, 510 (1979).
\bibitem{Mermin70}
  N.D.~Mermin,
  Phys. Rev. B {\bf 1}, 2362 (1970).
\bibitem{Reinholz00}
  H.~Reinholz et al.,
  Phys. Rev. E {\bf 62}, 5648 (2000).
\bibitem{Seuferling89}
  P.~Seuferling et al.,
  Phys. Rev. A{\bf 40}, 323 (1989).
\bibitem{Arkhipov98}
  Y.V.~Arkhipov, A.E.~Davletov,
  Phys. Lett. A {\bf 247}, 339 (1998).
\bibitem{Gregori06}
  G.~Gregori, et al.,
  Phys. Rev. E {\bf 74}, 026402 (2006).
\bibitem{Schwarz06}
  V.~Schwarz, et al.,
  {\it Hypernetted chain calculations for two--component plasmas},
  Cont. Plasma Phys (2006) submitted.
\bibitem{Sitenko95}
  A.~Sitenko, V.~Malnev,
  {\it Plasma Physics Theory},
  Chapman \& Hall, London (1995).
\bibitem{Feldhaus97}
  J.~Feldhaus, et al.,
  Opt. Commun. {\bf 140}, 341 (1997).
\bibitem{4gls}
\href{http://www.4gls.ac.uk}{http://www.4gls.ac.uk}
\bibitem{Gross49}
  D.~Bohm, E.P.~Gross, 
  Phys. Rev. {\bf 75}, 1851 (1949).
\bibitem{Zimmermann98}
  R.~Zimmermann,
  {\it Many--Particle Theory of Highly Excited Semiconductors},
  Teubner, Leipzig (1998).
\bibitem{Kraeft86}
  W.-D.~Kraeft, et al.,
  {\it Quantum Statistics of Charged Particle Systems},
  Akademie-Verlag, Berlin (1986).
\bibitem{Kramers23} 
  H.A.~Kramers,
  Philos. Mag. {\bf 46}, 836 (1923).
\bibitem{Rybi75} 
  G.B.~Rybicki, A.P.~Lightman, 
  {\it Radiative Processes in Astrophysics}, 
  J. Wiley \& Sons, New York (1975).
\bibitem{Gaunt30} 
  J.A.~Gaunt,
  Proc. R. Soc. A {\bf 126}, 654 (1930).
\bibitem{Fortmann06} 
  C.~Fortmann, et al.,
  High Energy Density Physics {\bf 2}, 57 (2006).
\bibitem{Meyer05}
  M.~Meyer, et al.,
  Phys. Rev A {\bf 74}, 011401(R) (2006).
\bibitem{Berden04} 
  G.~Berden, et al.,
   Phys. Rev. Lett. {\bf 93}, 114802 (2004).
\bibitem{Slipchenko02}  
  M.N.~Slipchenko, et al.,
  Rev. Sci. Instrum. {\bf 73}, 3600 (2002).
\bibitem{Grams05}
  M.~Grams, et al.,
   Rev. Sci. Instrum. {\bf 76}, 123904 (2005).
\bibitem{Trostell95}
  B.~Trostell,
   Nucl. Instr. Meth. in Phys. Res. A {\bf 362}, 41 (1995).
\bibitem{Jasny94} 
  J.~Jasny, et al.,
  Rev. Sci. Instrum. {\bf 65}, 1631 (1994).
\bibitem{Hoell06} 
  A.~H\"oll, et al.,
  {\it Thomson Scattering Measurements of Plasma Dynamics},
  FLASH proposal (2006).
\bibitem{Zubarev97}
  D.N.~Zubarev, V.~Morozov, G.~R\"opke,
  {\it Statistical Mechanics of Nonequilibrium  Processes},
  Akademie Verlag, Berlin (1997).
\end{thebibliography}
\end{document}